\documentclass[12pt]{article}
\usepackage{amsmath,amssymb,graphicx}
\usepackage{geometry}
\usepackage{slashed}
\usepackage[numbers]{natbib}
\usepackage{float}
\usepackage[colorlinks=True, citecolor=blue,linkcolor=blue, urlcolor=blue]{hyperref}
\usepackage{url}
\usepackage{simplewick}
\usepackage{simpler-wick}
\usepackage{xcolor}
\usepackage[normalem]{ulem}
\definecolor{red}{rgb}{0.8,0,0}
\definecolor{violet}{rgb}{0.4,0,0.4}
\definecolor{green}{rgb}{0,0.5,0.0}
\definecolor{navy}{rgb}{0.0,0.0,0.6}
\definecolor{orange}{rgb}{0.8,0.2,0.0}

\usepackage[normalem]{ulem}

\title{Neutron star cooling with $\Delta$ dUrca processes}

\author{Yogeesh N, Monika Sinha, Vishal Parmar }

\date{\today}
             
\begin{document}
\maketitle
\begin{abstract}
The cooling of neutron stars (NSs) provides insight into their composition and equation of state. However, a complete understanding of this phenomenon is still an open question. The early stage of NS cooling is dominated by neutrino emission from the core, which rapidly cools the star. Urca processes are one such mechanism where neutrinos escape from the core, carrying away most of the heat. Nucleonic and hyperonic direct Urca processes are known rapid cooling mechanisms. The possible appearance of $\Delta$ baryons in dense matter introduces additional weak-interaction channels, including $\Delta$ direct Urca processes. Although such processes have been discussed in the context of $\Delta$-admixed NS matter, their neutrino emissivities and impact on NS cooling have not been systematically investigated. In this work, for the first time we incorporate the neutrino emissivities of the relevant $\Delta$ direct Urca processes into NS cooling simulations. We investigate the thermal evolution of $\Delta$-admixed NSs and assess the impact of these additional neutrino emission channels on their cooling behavior. We find that $\Delta$ direct Urca processes can provide an efficient neutrino-cooling mechanism and significantly accelerate the thermal evolution of NSs. Our results demonstrate that $\Delta$ baryons and their associated weak-interaction processes can play an important role in explaining NSs with unusually low surface temperatures and provide a new connection between the composition of dense matter and NS cooling observations.
\end{abstract}

\maketitle

\section{Introduction \label{Introduction} }
Neutron star (NS) cooling remains one of the important open problems in NS astrophysics, primarily because the composition and equation of state (EOS) of matter at supranuclear densities are still not well constrained. A newly born NS, formed shortly after a core-collapse supernova, is initially extremely hot, with an internal temperature of the order of $10^{10}$--$10^{11}$ K \cite{Potekhin2015}. During the first several seconds to hundreds of seconds, the proto-NS undergoes rapid deleptonization and cooling through the emission of neutrinos. As the star evolves toward a thermally relaxed configuration, its surface temperature decreases to approximately $10^{8}$--$10^{9}$ K.  During the early and intermediate stages of its evolution, neutrino emission from the stellar core is the dominant cooling mechanism. Several neutrino emission processes can contribute, including the modified Urca (mUrca) process, nucleon--nucleon bremsstrahlung etc. These are slow neutrino emitting process and therefore lead to slow cooling of NSs. In contrast, the comparatively fast neutrino-emitting process, direct Urca (dURCA), occurs at sufficiently high densities in the inner regions of massive neutron stars, resulting in rapid cooling. The mURCA operates throughout the star while dURCA operates in inner core of massive stars only. Consequently, the contribution of mURCA to the overall cooling rate can be subdominant when dURCA is kinematically allowed.

However, in massive neutron stars, the appearance of exotic degrees of freedom in the inner core can open additional dUrca-like neutrino-emission channels. These processes can provide efficient pathways for neutrino cooling and thereby make significant contributions to the overall cooling rate of the star. For example, the appearance of hyperons, antikaon condensates, or deconfined strange quark matter can give rise to additional dUrca-like reactions \cite{Dapo2010,Vidana2016,Malik2022,Bauswein2026,Haensel1981,Glendenning1998kaon,Parmar2024,Shapiro1983,Glendenning1985,Kumar2023}, which can substantially modify the thermal evolution of neutron stars. Observationally, some slowly cooling neutron stars are consistent with cooling models in which the  mUrcA process provides the dominant neutrino-emission mechanism. In contrast, some neutron stars exhibit comparatively rapid cooling that may require additional dUrca-like processes. Therefore, comparing the observed thermal evolution of neutron stars with theoretically calculated cooling curves for different compositions of dense matter in the inner core can provide a valuable test of models of neutron-star interior matter.

An important aspect of the composition of dense matter is the possible appearance of baryonic resonances, particularly the $\Delta$ baryons. The $\Delta$ baryons constitute the lowest lying baryonic resonances and may appear in the core of NSs at densities several times the nuclear saturation density ($\rho_0$). The presence of $\Delta$ baryons can modify the composition of NS matter, the particle fractions, the EOS, and consequently the macroscopic properties of NSs such as their mass--radius relation and maximum mass \cite{Parmar2024}. The inclusion of $\Delta$ baryons can also have a significant impact on the conventional nucleonic dUrca (NdUrca) processes. Ref. \cite{JiaJie2018} showed that the appearance of $\Delta$ baryons modifies the particle composition of NS matter and can lead to the opening of the dUrca channels comparatively low densities, allowing these processes to operate over a larger fraction of the NS. 
In addition, the emergence of $\Delta$ baryons provides new channels for weak interactions that can contribute to the neutrino emissivity of the stellar core \cite{Prakash1992,Yakovlev2001}.

In particular, weak processes involving $\Delta$ baryons can operate through dUrca reactions with the general form
\begin{equation}
    \Delta \rightarrow B+e^-+\bar{\nu}_e, B+e^- \rightarrow \Delta + \nu_e
\end{equation}
where $\Delta$ generically denotes the $\Delta^-$ and $\Delta^0$ states and $B$ denotes $p$, $n$ and $\Lambda$.
 These reactions can provide an additional source of neutrino emission and may significantly influence the thermal evolution of NSs. Since dUrca processes have emissivities that are substantially larger than those of mUrca processes, the opening of a $\Delta$ dUrca channel may lead to enhanced cooling over the corresponding density range.

In this work, we investigate the impact of $\Delta$ baryons on the thermal evolution of NSs by incorporating $\Delta$ degrees of freedom into the EOS and consistently including the corresponding $\Delta$ dUrca neutrino emission processes in the NS cooling calculations.
The neutrino emissivities associated with the relevant $\Delta$ dUrca reactions are studied in ref \cite{N:2026psg} (Hereafter, paper-I). Consequently, we study their influence on the cooling rate of NSs. By comparing cooling curves obtained with and without the $\Delta$ dUrca contributions, we assess the extent to which these additional weak-interaction channels modify the thermal evolution. Using the emissivity derived in our previous work \cite{N:2026psg} and computing the cooling rate with the help of NSCool code \footnote{\url{https://www.astroscu.unam.mx/neutrones/NSCool/}} \cite{Page2013}, we examine its dependence on stellar mass, weak-interaction coupling parameters, and the composition of the stellar envelope. We compare the resulting cooling behavior with that obtained from the conventional N dUrca process.

\section{NS cooling in presence of $\Delta$ baryons}

The appearance of $\Delta$ baryons in NSs can significantly affect their properties, including the softening of the EOS, modification of the proton fraction, and reduction of the threshold density for the NdUrca process \cite{Oliveira:2007zz,JiaJie2018,Ribes2019,Vishal2025}. To investigate these effects, we employ the relativistic mean-field (RMF) theory with the DD-ME2 parametrization. The matter is assumed to consist of neutrons ($n$), protons ($p$), electrons ($e^-$), $\Lambda$ hyperons, and $\Delta$ baryons, with the interactions
mediated by the exchange of $\sigma$, $\omega$, and $\rho$ mesons. The details of the model parameters and the corresponding coupling constants are provided in \cite{Vivek2021}.

Since the $\Delta$--meson couplings are not yet well constrained by experimental or observational data, we adopt the coupling ratios used in Ref.~\cite{Raduta2021}. In particular, we define $x_{\sigma\Delta}=g_{\sigma\Delta}/g_{\sigma N}$, $x_{\omega\Delta}=g_{\omega\Delta}/g_{\omega N}$, and $x_{\rho\Delta}=g_{\rho\Delta}/g_{\rho N}$, where $g_{\sigma\Delta}$, $g_{\omega\Delta}$, and $g_{\rho\Delta}$ denote the couplings of the $\Delta$ baryons to the $\sigma$, $\omega$, and $\rho$ mesons, respectively, while $g_{\sigma N}$, $g_{\omega N}$, and $g_{\rho N}$ are the corresponding nucleon couplings. We employ  $x_{\sigma\Delta}=1.18$, $x_{\omega\Delta}=1.0$, and $x_{\rho\Delta}=1.0$. The onset densities of the $\Delta$ dUrca processes and paritcles participating in these processes are given in Table \ref{tab:durca_thresholds}.

\begin{table*}[htbp]
\centering
\renewcommand{\arraystretch}{1.2}
\setlength{\tabcolsep}{4pt}
\begin{tabular}{cccccc}
\hline\hline
\multicolumn{6}{c}{Threshold densities of appearance ($n_0$)} \\
\hline
$\Delta^-$ & $\Delta^0$ & $\Lambda$ & 
\shortstack{$\Delta^- \rightarrow n + e^- + \bar{\nu}_e$} & 
\shortstack{$\Delta^0 \rightarrow p + e^- + \bar{\nu}_e$} & 
\shortstack{$\Delta^- \rightarrow \Lambda + e^- + \bar{\nu}_e$} \\
\hline
1.33 & 2.26 & 3.09 & 3.35 & 3.55 & 3.93 \\
\hline\hline
\end{tabular}
\caption{Threshold densities for the appearance of particles and the onset of $\Delta$ dUrca processes obtained using the DD-ME2 parameterization.}
\label{tab:durca_thresholds}
\end{table*}

\subsection{Emissivity}
The neutrino emissivity associated with the $\Delta$ dUrca processes was derived in our previous work~\cite{N:2026psg}. Here, we briefly summarize the relevant expression and focus on its implementation in the NS cooling. The $\Delta$ dUrca processes and their inverse processes are,

\begin{eqnarray}
    \Delta^- &\rightarrow& n + e^- + \bar{\nu}_e,
    \qquad
    n + e^- \rightarrow \Delta^- + \nu_e\\[4pt]
    \Delta^- &\rightarrow& \Lambda + e^- + \bar{\nu}_e,
    \qquad
    \Lambda + e^- \rightarrow \Delta^- + \nu_e\\[4pt]
    \Delta^0 &\rightarrow& p + e^- + \bar{\nu}_e,
    \qquad
    p + e^-\rightarrow \Delta^0 + \nu_e.
\end{eqnarray}

The $\Delta$ dUrca processes considered here corresponds to a charged-current
weak transition between a spin-$3/2$ $\Delta$ baryon and a spin-$1/2$ baryon,
accompanied by the emission of a charged lepton and an antineutrino. The
corresponding weak interaction is mediated by the charged weak boson and can
be described by the effective interaction Hamiltonian
\begin{equation}
\mathcal{H}_{\rm int}
=
\frac{g^2}{2}
\left\{
\bar{\psi}_{B}\Gamma^{\mu}W_{\mu}\psi_{\Delta,\sigma}
\right\}
\left\{
\bar{\psi}_{\ell}\gamma_{\nu}(1-\gamma_5)W^{\nu}\psi_{\nu}
\right\},
\label{eq:weak_hamiltonian}
\end{equation}
where $g$ is the weak coupling constant, $W_\mu$ denotes the charged weak
boson field, and $\Gamma^\mu$ represents the hadronic transition vertex
between the $\Delta$ and final baryon states. The spin-$3/2\rightarrow1/2$
structure of the weak transition is incorporated through the corresponding
transition vertex, which we write as \cite{Wang2022,Zhao2024},
\begin{equation}
     \Gamma^{\mu\sigma} = \gamma^{\mu}\Big[\frac{p^{\sigma}_{B}}{m_{B}}(f_1+g_1\,\gamma_5)\Big],
\end{equation}
where $f_1$ and $g_1$ denote the vector and axial-vector weak form factors,
respectively. Here, $p_B$ denotes the four-momentum of the final-state
baryon, and $m_B$ is its mass. This interaction Hamiltonian provides the starting point for
evaluating the matrix element and, subsequently, the neutrino emissivity of
the $\Delta$ dUrca processes.

For a degenerate NS core, the neutrino emissivity for the process $\Delta \rightarrow B+e^-+\bar{\nu}_e$ and its inverse, where $B$ denotes a baryon, is given by
\begin{align} Q &= \frac{457\pi G_F^2 C^2}{3780\, m_\Delta^2 m_B^2} \, T_9^6 \Big[(m_\Delta^2 - m_B^2)^2 - 2 m_e^2 (m_\Delta^2 + m_B^2)\Big] \nonumber \\ &\quad \times \left(f_1^2 + 3 g_1^2\right) \, m^*_\Delta m^*_B m^*_e \, \Theta_{\Delta B e}, \label{eq:Q} \end{align}
where $G_F$ is the Fermi weak coupling constant, $C$ is the appropriate Cabibbo factor, and $m_{\Delta}$ and $m_B$ denote the vacuum masses of the $\Delta$ isobar and baryon $B$, respectively. The quantities $m_\Delta^*$ and $m_B^*$ are the corresponding effective masses in dense matter, while $m_e^*=\mu_e/c^2$, with $\mu_e$ being the electron chemical potential and $c$ the speed of light in vacuum. Here, $T_9=T/(10^9~\mathrm{K})$ denotes the temperature in units of $10^9$ K. The threshold function $\Theta_{\Delta B e}$ enforces the kinematic condition for the occurrence of the $\Delta$ dUrca process and vanishes when the corresponding momentum-conservation condition is not satisfied. The complete derivation of Eq.~\eqref{eq:Q} and the expressions for the weak transition matrix elements are given in paper-I.

\subsection{Implementation in NSCool}
We investigate the thermal evolution of neutron stars using the publicly available one-dimensional NS cooling code \texttt{NSCool} written in Fortran 77 \cite{Page2013}. The code assumes spherical symmetry and solves the general-relativistic equations of energy balance and heat transport throughout the stellar interior, accounting for the relevant contributions from the specific heat, thermal conductivity, neutrino emissivity, and surface photon luminosity.
 The local energy balance is governed by the specific heat
and the neutrino emissivity. In the general-relativistic formulation, the
energy-balance equation can be written as \footnote{\url{https://www.astroscu.unam.mx/neutrones/NSCool/NSCool_Lecture_1.pdf}}
\begin{equation}
\frac{d}{dr}\left(L e^{2\phi}\right)
=
-\frac{4\pi r^2 e^{\phi}}
{\sqrt{1-2Gm(r)/(rc^2)}}
\left[
\frac{dU}{dt}
+e^{\phi}\left(Q_\nu-Q_h\right)
\right],
\label{eq:energy_balance}
\end{equation}
where $L$ is the local luminosity, $\phi$ is the gravitational redshift
potential, $m(r)$ is the gravitational mass enclosed within radius $r$,
$Q_\nu$ is the neutrino emissivity, and $Q_h$ represents any local heating
rate. The change in internal energy can be expressed in terms of the specific
heat as
\begin{equation}
\frac{dU}{dt}=C_V\frac{dT}{dt},
\label{eq:internal_energy}
\end{equation}
where $C_V$ is the specific heat per unit volume and $T$ is the local
temperature. In the present work, the total neutrino emissivity entering
Eq.~\eqref{eq:energy_balance} is augmented by the $\Delta$ dUrca
contribution derived in this work,
\begin{equation}
Q_\nu = Q_\nu^{\rm standard}+Q_\nu^{\Delta}.
\label{eq:total_emissivity}
\end{equation}
The $\Delta$ dUrca contribution is evaluated locally and included only in
regions where the corresponding kinematic conditions are satisfied.

\texttt{NSCool} includes a variety of standard neutrino-emission processes, such as the mUrca, dUrca, nucleon--nucleon bremsstrahlung, and Cooper-pair breaking and formation processes, together with the corresponding effects of baryon superfluidity. We extend the standard \texttt{NSCool} framework to incorporate the $\Delta$ dUrca process as an additional neutrino-emission channel. Specifically, we implement the derived $\Delta$ dUrca emissivity as a local neutrino-emission rate in the core and activate it only in regions where the corresponding kinematic conditions for the process are satisfied. The emissivity is evaluated using the local thermodynamic quantities, effective masses, particle fractions, and chemical potentials obtained from the adopted EOS. The resulting $\Delta$ dUrca contribution is then included in the total neutrino emissivity used by \texttt{NSCool} to evolve the internal temperature and surface temperature as functions of time. This modification allows us to consistently assess the impact of the $\Delta$ dUrca process on the thermal evolution of NSs over a range of stellar masses and model parameters. 

The resulting internal temperature is related to the observable effective surface temperature through the corresponding envelope model. We consider both iron (Fe) and light-element (H) envelopes to investigate the sensitivity of the observable cooling curves to the envelope composition. These calculations allow us to separately assess the effects of stellar mass, weak-interaction uncertainties, and envelope composition on the cooling signatures associated with the $\Delta$ dUrca process.

\section{Results}

\begin{figure}[h!]
    \centering
    \includegraphics[width=1.0\linewidth]{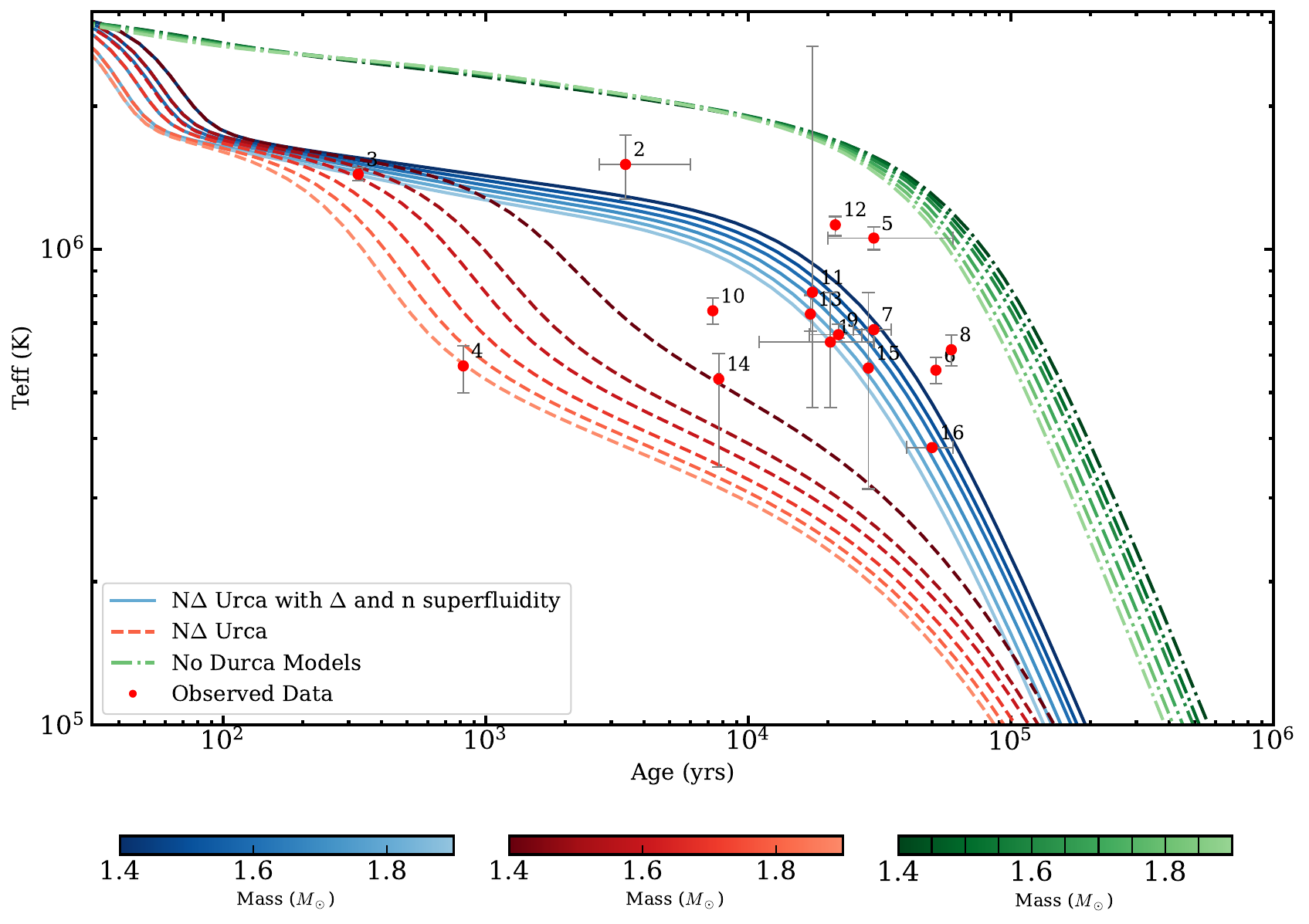}
    \caption{Cooling evolution of NSs with masses in the range
$1.4\leq M/M_\odot\leq1.9$. The observational data has been taken from the Ref. \cite{Potekhin2020}.}
    \label{fig:Emissivity}
\end{figure}

We investigate the impact of the $\Delta$ dUrca process on the thermal evolution of NSs by implementing the emissivity derived in paper-I in the \texttt{NSCool} code. We consider stellar masses in the range $1.4\leq M/M_\odot\leq1.9$ and compare the resulting cooling curves with the observed surface temperatures of isolated NSs. The observational data used in the comparison are shown as red points in Figs.~\ref{fig:Emissivity}--\ref{fig:envelope}. Since the $\Delta$ pairing gap in dense matter is not yet well constrained microscopically, we adopt a representative value of 1 MeV for the calculations including $\Delta$ superfluidity. This choice is intended to investigate the possible impact of $\Delta$ pairing on the cooling evolution rather than to represent a uniquely determined value of the gap.

Fig.~\ref{fig:Emissivity} compares three sets of cooling sequences: models without a dUrca process, models including the N$\Delta$ dUrca process, and models including the N$\Delta$ dUrca process together with neutron and $\Delta$ superfluidity. The stellar masses considered range from $1.4$ to $1.9\,M_\odot$, with the corresponding masses indicated by the color scales. The inclusion of the $\Delta$ dUrca process substantially enhances the neutrino luminosity and leads to faster thermal evolution compared with models without a dUrca process. Consequently, the N$\Delta$ dUrca sequences extend toward lower surface temperatures and populate a region of the age--temperature diagram that is not reached by the corresponding no-dUrca models.

The effect is particularly relevant for the relatively young and cold objects in the observational sample. Among these sources, PSR J0205+6449 (4) is a notable example, with an age of approximately $8.2\times10^{2}$ yr and an effective temperature of $\sim 5\times10^{5}$ K. The high-mass N$\Delta$ dUrca sequences reach the region occupied by this source, with the $M\simeq1.9\,M_\odot$ sequence providing the closest correspondence. Similarly, PSR B2334+61 (14), with an age of approximately $7.7\times10^{3}$ yr and an effective temperature of $\sim4.6\times10^{5}$ K, lies close to the high-mass N$\Delta$ dUrca sequences. The $M\simeq1.8$--$1.9\,M_\odot$ models provide the closest correspondence to this source. These results indicate that sufficiently massive NSs undergoing N$\Delta$ dUrca cooling can naturally reach the relatively low temperatures observed in young NSs.

Other relatively cool sources are also located close to the N$\Delta$ dUrca cooling curves. For example, PSR B1727-47 (16), with an age of approximately $5\times10^{4}$ yr and an effective temperature of $\sim3.3\times10^{5}$ K, lies near the lower-temperature part of the N$\Delta$ dUrca sequences. Likewise, PSR J0837-2454 (15), with an age of $\sim2.9\times10^{4}$ yr and an effective temperature of $\sim4.9\times10^{5}$ K, is located in the region covered by the N$\Delta$ dUrca models. Thus, the inclusion of the N$\Delta$ dUrca process considerably enlarges the range of surface temperatures accessible during the first $10^4$--$10^5$ yr of NS evolution.

\begin{table*}[t]
\centering
\renewcommand{\arraystretch}{1.2}
\setlength{\tabcolsep}{8pt}

\label{tab:observed_sources}

\begin{tabular}{c l c c}
\hline
\hline
No. & Source & Age (yr) & $T_{\rm s}$ (K) \\
\hline
1  & 2XMM J104608.7$-$594306 
   & $2.05^{+0.95}_{-0.95}\times10^{4}$ 
   & $(6.38^{+1.74}_{-1.74})\times10^{5}$ \\

2  & CXOU J181852.0$-$150213 
   & $3.40^{+2.60}_{-0.70}\times10^{3}$ 
   & $(1.51^{+0.23}_{-0.23})\times10^{6}$ \\

3  & CXOU J232327.8+584842 
   & $326.2^{+12.5}_{-6.7}$ 
   & $(1.44^{+0.05}_{-0.05})\times10^{6}$ \\

4  & PSR J0205+6449 
   & $819.5^{+0.1}_{-0.1}$ 
   & $(5.69^{+0.58}_{-0.70})\times10^{5}$ \\

5  & PSR J0538+2817 
   & $3.00^{+3.00}_{-1.00}\times10^{4}$ 
   & $(1.06^{+0.06}_{-0.06})\times10^{6}$ \\

6  & PSR J0554+3107 
   & $5.17\times10^{4}$ 
   & $(5.57^{+0.35}_{-0.35})\times10^{5}$ \\

7  & CXOU J061705.3+222127 
   & $3.00^{+0.50}_{-0.50}\times10^{4}$ 
   & $(6.78^{+0.07}_{-0.05})\times10^{5}$ \\

8  & PSR J0633+0632 
   & $5.92\times10^{4}$ 
   & $(6.15^{+0.46}_{-0.46})\times10^{5}$ \\

9  & PSR B0833$-$45 
   & $2.20^{+0.50}_{-0.50}\times10^{4}$ 
   & $(6.61^{+0.35}_{-0.12})\times10^{5}$ \\

10 & PSR J1357$-$6429 
   & $7.31\times10^{3}$ 
   & $(7.43^{+0.46}_{-0.46})\times10^{5}$ \\

11 & PSR B1706$-$44 
   & $1.75\times10^{4}$ 
   & $(8.12^{+18.57}_{-3.48})\times10^{5}$ \\

12 & PSR B1823$-$13 
   & $2.14\times10^{4}$ 
   & $(1.13^{+0.05}_{-0.06})\times10^{6}$ \\

13 & PSR J2021+3651 
   & $1.72\times10^{4}$ 
   & $(7.31^{+0.70}_{-0.58})\times10^{5}$ \\

14 & PSR B2334+61 
   & $7.70^{+0.10}_{-0.10}\times10^{3}$ 
   & $(5.34^{+0.70}_{-1.86})\times10^{5}$ \\

15 & PSR J0837$-$2454 
   & $2.86\times10^{4}$ 
   & $(5.63^{+2.50}_{-2.50})\times10^{5}$ \\

16 & PSR B1727$-$47 
   & $5.00^{+1.00}_{-1.00}\times10^{4}$ 
   & $3.83\times10^{5}$ \\

\hline
\hline
\end{tabular}
\caption{Observed NS sources used for comparison with the cooling curves.
The ages and effective surface temperatures are given with their corresponding
lower and upper limits.}
\end{table*}

The effect of $\Delta$ superfluidity is illustrated by the corresponding cooling sequences that include both neutron and $\Delta$ pairing. The presence of $\Delta$ superfluidity modifies the cooling evolution by reducing the contribution of $\Delta$ baryons to the relevant phase space below the corresponding critical temperature. As a result, the cooling curves differ from those obtained with the N$\Delta$ dUrca process in the absence of $\Delta$ pairing. The resulting sequences can provide a closer correspondence to some of the relatively cold observational data. Since the magnitude and density dependence of the $\Delta$ pairing gap remain poorly constrained, however, the quantitative location of these cooling curves should be regarded as model dependent. Our choice of 1 MeV therefore serves as a representative benchmark for assessing the possible role of $\Delta$ superfluidity.

The mass dependence of the cooling curves is also important. For the adopted EOS, increasing the stellar mass increases the central density and consequently enhances the role of the $\Delta$ dUrca process once the corresponding threshold is reached. The higher-mass sequences therefore undergo more efficient neutrino cooling and populate the colder region of the age--temperature diagram. In particular, the $M\simeq1.8$--$1.9\,M_\odot$ sequences provide the closest correspondence to the young and relatively cold objects, whereas lower-mass sequences remain comparatively warmer.

\subsection{Dependence on the meson--$\Delta$ couplings}
\label{subsec:meson_couplings}

The properties of $\Delta$ baryons in dense matter are subject to
considerable uncertainty, in particular because the couplings between
the $\Delta$ baryons and the mesonic fields are not well constrained.
These couplings determine the appearance and abundance of $\Delta$
baryons in the stellar core and consequently influence the density
range over which the $\Delta$ dUrca process can operate. It is
therefore important to examine the sensitivity of the cooling
predictions to variations in the meson--$\Delta$ coupling parameters.

\begin{figure}[h!]
    \centering
    \includegraphics[width=1.0\linewidth]{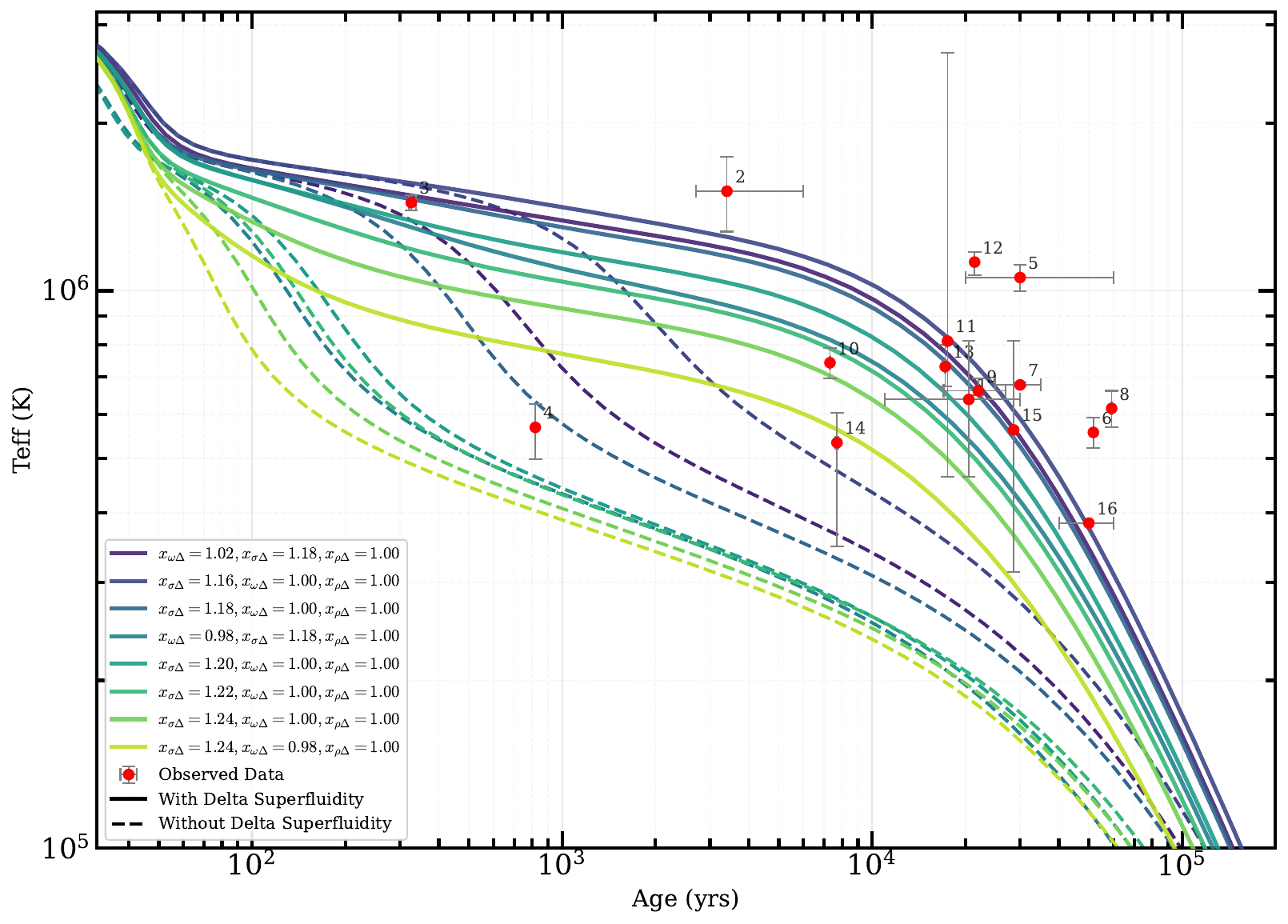}
    \caption{Sensitivity of the NS cooling evolution to the
meson--$\Delta$ coupling parameters of a $1.8$ $M_\odot$ star. The different cooling sequences
correspond to the variations of the scaling parameters $x_{\omega\Delta}$,
$x_{\sigma\Delta}$, and $x_{\rho\Delta}$ considered in this work. Solid curves represent
models including $\Delta$ superfluidity, whereas dashed curves
correspond to models without $\Delta$ superfluidity}
    \label{fig:meson_couplings}
\end{figure}

Fig.~\ref{fig:meson_couplings} shows the cooling sequences obtained by varying
the meson--$\Delta$ coupling parameters over the range considered in
this work. The different curves correspond to different combinations
of the scaling parameters $x_{\omega\Delta}$,
$x_{\sigma\Delta}$, and $x_{\rho\Delta}$, which
characterize the corresponding variations in the meson--$\Delta$
couplings. The solid curves represent models including $\Delta$
superfluidity, while the dashed curves correspond to models without
$\Delta$ superfluidity. The stellar masses span the range
$1.4$--$1.9\,M_\odot$.

The comparison between the solid and dashed curves further illustrates
the role of $\Delta$ superfluidity. Including $\Delta$ pairing changes
the cooling evolution relative to the corresponding models without
$\Delta$ superfluidity, demonstrating that both the interaction
parameters and the pairing properties of $\Delta$ baryons can influence
the observable thermal evolution. 

Overall, the variation in the meson--$\Delta$ couplings further demonstrates the enhanced cooling associated with the $\Delta$ dUrca process across a broad range of coupling strengths. The persistence of relatively cold cooling
sequences over the range of coupling parameters considered here
supports the possibility that N$\Delta$ dUrca emission can play an
important role in the thermal evolution of sufficiently massive
NSs.

\subsection{Dependence on the envelope composition}

\begin{figure}[htbp]
    \centering
    \includegraphics[width=1\linewidth]{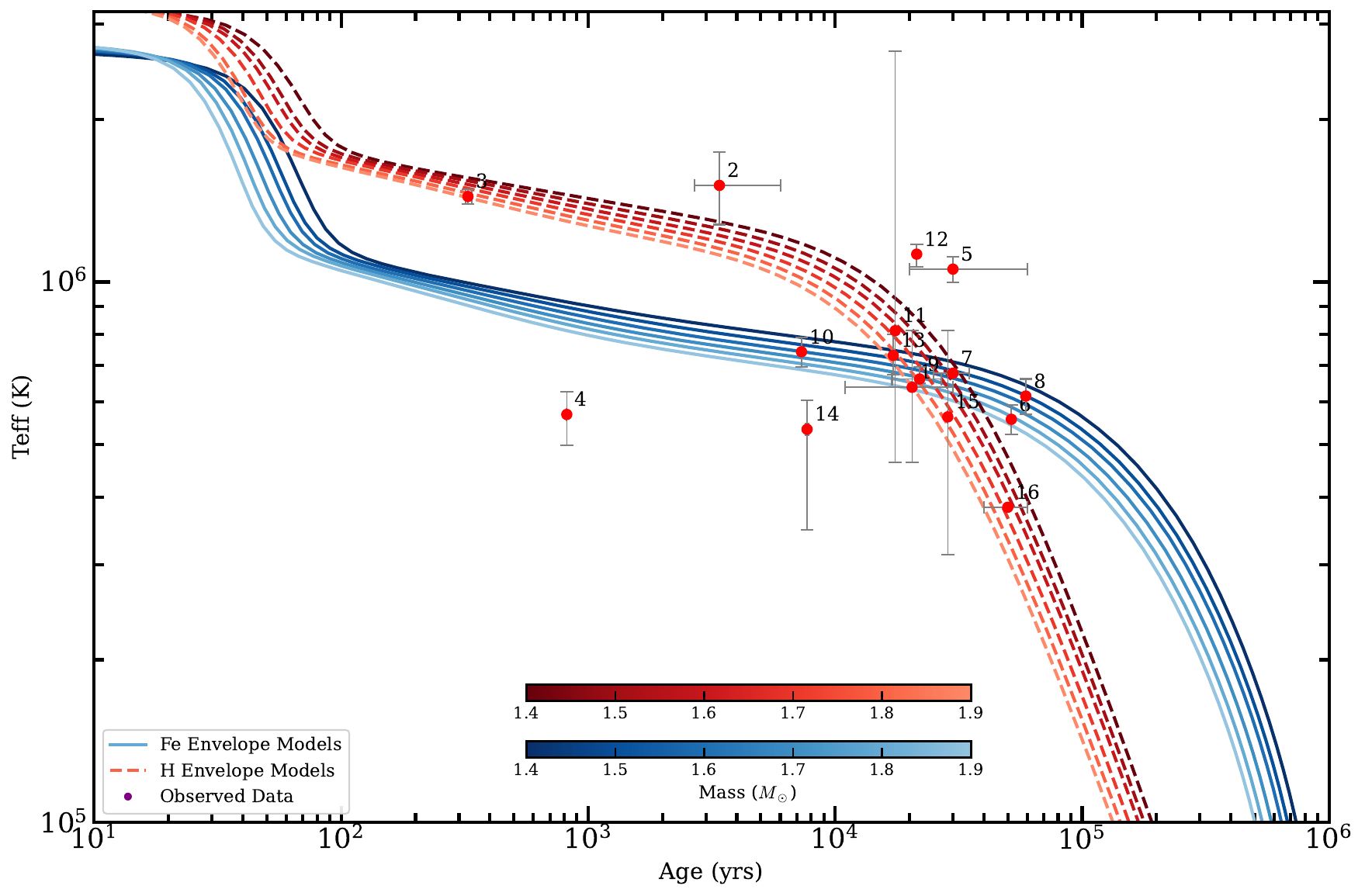}
    \caption{Dependence of the NS cooling evolution on the
composition of the stellar envelope.The solid blue curves correspond
to models with an iron (Fe) envelope, while the dashed red curves
represent models with a hydrogen (H) envelope. }
    \label{fig:envelope}
\end{figure}

We finally examine the dependence of the cooling evolution on the assumed composition of the NS envelope. Fig.~\ref{fig:envelope} compares the cooling sequences obtained with iron (Fe) and hydrogen (H) envelopes for stellar masses in the range $1.4$--$1.9\,M_\odot$.

The envelope composition determines the relation between the temperature of the stellar interior and the observable effective surface temperature. Consequently, Fe- and H-envelope models with otherwise identical interior physics can exhibit different surface-temperature evolution. As shown in Fig.~\ref{fig:envelope}, the H-envelope models generally yield higher effective surface temperatures than the corresponding Fe-envelope models at a given stage of the thermal evolution. The difference is particularly relevant during the early and intermediate stages of cooling.

Importantly, the characteristic effect of the N$\Delta$ dUrca process persists for both envelope prescriptions. Although the envelope composition changes the detailed location of the cooling curves in the age--temperature plane, the high-mass N$\Delta$ dUrca sequences continue to extend toward the relatively cold observational region. 
Thus, the enhanced cooling associated with the N$\Delta$ dUrca process is not a consequence of adopting a particular envelope composition.

\subsection{Implications for NS cooling}

The results presented above demonstrate the potential importance of the N$\Delta$ dUrca process for NS thermal evolution. To our knowledge, this is the first cooling study in which the $\Delta$-induced dUrca emissivity derived in this work is incorporated into a NS cooling calculation. The resulting sequences show that the $\Delta$ dUrca processes can provide an efficient neutrino-cooling channel capable of producing substantially colder stars than models without the corresponding process.

The effect is most pronounced for the higher-mass stars, for which the central density is sufficiently high for the $\Delta$ dUrca process to operate. The $M\simeq1.8$--$1.9\,M_\odot$ sequences provide the closest correspondence to the relatively young and cold sources PSR J0205+6449 (4) and PSR B2334+61 (14). In particular, the $M\simeq1.9\,M_\odot$ sequence passes closest to PSR J0205+6449 (4), while the $M\simeq1.8$--$1.9\,M_\odot$ sequences approach PSR B2334+61 (14). These comparisons suggest that $\Delta$-induced dUrca emission can provide a possible explanation for the relatively low surface temperatures of young NSs.

\section{Conclusions}

In this work, we have investigated the role of $\Delta$ baryons in the thermal evolution of NSs by incorporating the emissivity of the $\Delta$ dUrca process into the publicly available \texttt{NSCool} code. We have examined the resulting cooling curves for NS masses in the range $1.4$--$1.9\,M_\odot$ and compared them with the observed surface temperatures of isolated NSs.

Our results demonstrate that the $\Delta$ dUrca process can provide an efficient neutrino-cooling mechanism once the stellar density is sufficiently high for the process to operate. Compared with models without a dUrca process, the inclusion of $\Delta$ dUrca leads to substantially faster cooling and allows the theoretical trajectories to extend into the relatively cold region of the observed age--temperature diagram. The effect is particularly pronounced for the higher-mass stars, for which the central density is higher and the $\Delta$ dUrca process can contribute over a larger volume of the stellar core.

The comparison with observations shows that the high-mass sequences, particularly those with $M\simeq1.8$--$1.9\,M_\odot$, can approach the locations of several relatively young and cold NSs. In particular, the $M\simeq1.9\,M_\odot$ sequence provides the closest correspondence to PSR J0205+6449 (4), while the $M\simeq1.8$--$1.9\,M_\odot$ sequences approach PSR B2334+61 (14). The cooling curves also pass through the region occupied by several relatively cold sources, including 2XMM J104608.7-594306 (1), PSR J0554+3107 (6), CXOU J061705.3+222127 (7), PSR B0833-45 (9), PSR B1706-44 (11), PSR J2021+3651 (13), PSR J0837-2454 (15), and PSR B1727-47 (16). These results indicate that the $\Delta$ dUrca process can provide a possible explanation for the relatively low surface temperatures observed in a subset of young NSs.

We have also investigated the effect of $\Delta$ superfluidity. Since the $\Delta$ pairing gap in dense matter is presently not well established, we have adopted a representative value of superfluid gap as 1 MeV. The inclusion of $\Delta$ superfluidity modifies the cooling curves and can lead to improved correspondence with some of the relatively cold observational data. However, the quantitative predictions associated with $\Delta$ pairing remain uncertain because both the magnitude and density dependence of the $\Delta$ pairing gap are poorly constrained. A more systematic treatment of microscopic $\Delta$ pairing gaps will therefore be important for future studies.

We have further examined the dependence of the results on the envelope composition by comparing Fe- and H-envelope models. As expected, the envelope composition modifies the relation between the internal and observable surface temperatures and therefore changes the detailed position of the cooling curves. Nevertheless, the enhanced cooling associated with the $\Delta$ dUrca process persists for both envelope prescriptions. This indicates that the main qualitative result is not strongly dependent on the assumed envelope composition.

Overall, our calculations demonstrate that $\Delta$-induced dUrca emission can represent an important additional fast neutrino-cooling mechanism in NSs containing $\Delta$ baryons. The ability of the $\Delta$ dUrca process to generate cooling curves reaching the relatively cold region occupied by young NSs provides a new possibility for interpreting the observed diversity of NS thermal evolution. A more quantitative assessment will require improved constraints on the $\Delta$ weak couplings, microscopic calculations of the $\Delta$ pairing gap, and a systematic exploration of the density dependence of $\Delta$ superfluidity. These developments will allow the role of $\Delta$ baryons in NS cooling to be tested more rigorously against future observations.

\bibliography{references}
\bibliographystyle{unsrtnat}
\end{document}